\documentstyle[preprint,aps]{revtex}

\begin{document}

\draft
\preprint{\parbox{5cm}{ADP-93-214/T132 \\
                       AUGUST 1993\\[4cm]
                       \strut}}
\title{Nuclear Shadowing at Small $x$ and $Q^2$.}

\author{W.Melnitchouk
        \footnote{Present address:
                  Institut f\"{u}r Theoretische Physik,
                  Universit\"{a}t Regensburg,
                  \mbox{D-93040} Regensburg, Germany.}
        and A.W.Thomas}
\address{Department of Physics and Mathematical Physics,
         University of Adelaide,        \\
         South Australia, 5005, Australia}

\maketitle

\begin{abstract}
Shadowing corrections to the structure functions of heavy nuclei are
calculated at very low values of Bjorken-$x$ and at values of the
momentum transfer relevant to recent experiments.
Good agreement is obtained with data from the E665 Collaboration
for $Xe/D$ and $Pb/D$, and with the NMC data on $Ca/D$ and $C/D$
structure function ratios.
Corrections to the deuteron structure function are also estimated
down to $x \sim 10^{-5}$, and found to be less than about $3\%$ over
the range of $x$ covered by the E665 data.
\end{abstract}



Recent experiments \cite{e665Xe,e665D} by the E665 Collaboration at Fermilab
have measured deep inelastic nuclear structure functions at values of
Bjorken-$x$ some two orders of magnitude smaller than
those attainable in previous experiments.
The data on heavy nuclei ($Xe$ and $Pb$) complement the existing
high-precision NMC data on lighter nuclei, such as
$Ca$ and $C$, which will help us to understand the origin of the
nuclear EMC effect at small $x$.
In addition, the E665 Collaboration measured the deuteron to proton
structure function ratio \cite{e665D}, from which the neutron structure
function was determined down to $x \sim 10^{-5}$.
To date, however, the shadowing corrections to $F_{2n}$ have not been
estimated in the region of $x$ and $Q^2$ covered by the E665 data.
We do so here.
Furthermore, we calculate the shadowing effect in heavier nuclei, which will
also serve as a check of the reliability of the deuteron predictions.

We consider a two-phase model which incorporates hadron and parton degrees of
freedom at low and high $Q^2$ values, respectively.
The approach is to use the vector meson dominance (VMD) model,
which describes low-$Q^2$, virtual-photon interactions,
and to model the approximate scaling of shadowing by Pomeron-exchange.
A synthesis of these two mechanisms is quite sensible if we are to investigate
the transition region between small- and large-$Q^2$ processes.
The physical picture is that the virtual photon interacts with the
nuclear target via its fluctuations into $q \bar q$ pairs.
If the virtuality of the photon is large the fluctuation is short-lived,
and a description in terms of diffractive scattering from partons is
appropriate.
If the virtuality is smaller, the virtual $q \bar q$ pair will have time
to evolve into a state resembling a vector meson, which then enables
a VMD-based description to be used.
In this sense the approach is similar to the model of shadowing adopted by
Kwiecinski \& Badelek \cite{kwbd}, and to the recent description of
real photon interactions by Schuler \& Sj\"{o}strand \cite{schsj}.

The empirical basis for the VMD model is the observation that
photon---hadron processes have many remarkable similarities with
purely hadronic reactions.
The most simple and natural explanation of this phenomenon is that the
photon itself has a hadronic (vector meson) structure.
Since VMD gives a very good description of shadowing in high energy
photoproduction, this suggests that the same mechanism may also be
responsible for shadowing in deep inelastic scattering (DIS),
at least at low $Q^2$.
Formally, the amount of shadowing can be quantified via the Glauber
multiple scattering expansion, which, in the eikonal approximation,
gives a correction to the nuclear structure function (per nucleon):
\begin{mathletters}
\label{VMD}
\begin{eqnarray}
A\ \delta^{(V)} F_{2A}(x,Q^2)
&=& { Q^2 \over \pi }
\sum_{V=\rho^0,\omega,\phi}
{ M_V^4\ \delta\sigma_{VA} \over f_V^2 (Q^2 + M_V^2)^2 }        \label{dFAV}
\end{eqnarray}
where
\begin{eqnarray}
\delta \sigma_{VA}
&=& -{ A (A-1)\ \sigma_{VN}^2 \over 2 }\
    {\rm Re}
    \int_{z'>z} d^2{\bf b}\ dz\ dz'\
    e^{i k_L (z'-z)}                                            \nonumber\\
& & \hspace*{0cm} \times\
    \rho^{(2)}({\bf b},z,z')\ \
    \exp \left( - {A\over 2} \int_{z}^{z'} {d\zeta \over L_V}
         \right)                                                \label{dsigVA}
\end{eqnarray}
is the shadowing correction to the $V A$ cross section, with
${\bf b}$ the impact parameter, $k_L^2 = M^2 x^2 (1 + M_V^2/Q^2)^2$,
and $M$ the nucleon mass.
\end{mathletters}
By the Heisenberg uncertainty principle only the lowest mass vector mesons
($\rho^0, \omega, \phi$) will be important, and these couple to the
photon with strengths $f_V$ = 2.28, 26.14 and 14.91,
respectively \cite{bauer}.
The approximation in Eqs.(\ref{VMD}) also omits non-diagonal vector meson
transitions ($VN \rightarrow V'N$), however these are not expected to be large.

In Eq.(\ref{dsigVA})
$\rho^{(2)}({\bf r},{\bf r'})
= N_{\cal C}\ \rho({\bf r})\ \rho({\bf r'})\
    \left\{ 1 - {\cal C}({\bf r}-{\bf r'}) \right\}$
is the two-body density function (with $\rho({\bf r})$ the single body density)
normalised so that
$ \int d^3{\bf r}\ d^3{\bf r'}\ \rho^{(2)}({\bf r},{\bf r'})
= \int d^3{\bf r}\ \rho({\bf r})
= 1$.
The correlation function ${\cal C}({\bf r}-{\bf r'})$ takes into account
the short-range repulsion of the $NN$ force, and can be modelled by
\begin{eqnarray}
{\cal C}({\bf r}-{\bf r'}) &=&
\left( { 3 j_1(\kappa |{\bf r}-{\bf r'}|)  \over
         \kappa |{\bf r}-{\bf r'}| }
\right)^2                                                       \label{corr}
\end{eqnarray}
where $\kappa = 3.6$ fm$^{-1}$ is chosen to reproduce a `hole' in the
two-body density which is $\approx$~0.5~fm wide at 1/2 maximum density.
For the single particle density in heavy nuclei ($A \agt 16$) we use the
Woods-Saxon (or Fermi) density, while for light nuclei ($A \alt 16$) an
harmonic oscillator (shell model) form is more appropriate \cite{gramsul}.

For light nuclei, the dominant process is that involving the double
scattering of vector mesons from two nucleons.
Higher order terms (multiple rescattering) in the Glauber expansion
attenuate the incident flux of vector mesons as they traverse
the nucleus, which will be progressively more important as $A$ increases.
The magnitude of the attenuation is determined by the mean free path of the
vector meson in the nucleus,
$L_V = (\sigma_{VN}\ \ \rho({\bf b},\zeta))^{-1}$
\cite{kwiec}.
For the total $VN$ cross sections we use the energy dependent forms
from Ref.\cite{dlsig}:\
$ \sigma_{\rho^0 N}
= \sigma_{\omega N}
= 13.63\ s^{\epsilon} + 31.79\ s^{-\eta} $
and
$ \sigma_{\phi N} = 10.01\ s^{\epsilon} - 1.51\ s^{-\eta} $,
where $s = (p+q)^2$, with $p$ and $q$ the nucleon and photon four-momenta,
respectively.
The parameters $\epsilon \approx 0.08$ and $\eta \approx 0.45$ are taken
from Regge theory.

Numerically, there is a strong $Q^2$ dependence of $\delta^{(V)} F_{2A}$,
which peaks at around $Q^2 \sim 1$ GeV$^2$.
For $Q^2 \rightarrow 0$, $\delta^{(V)} F_{2A}$ disappears due to the
vanishing of the total $F_{2A}$.
Furthermore, since this is a higher twist effect (because of the vector meson
propagators in Eq.(\ref{dFAV}) ), shadowing in the VMD model dies off
quite rapidly between $Q^2 \sim 1$ and 10 GeV$^2$, so that for
$Q^2 \agt 10$ GeV$^2$ it is almost negligible.
In order to reproduce the observed persistence of shadowing in this region
\cite{nmc}, one could extend the model by including additional hadronic
states, or a high-mass $q \bar q$ continuum, as in generalised
vector meson dominance models \cite{GVMD}.
However, when discussing DIS phenomena at high $Q^2$
it may be more efficient to use a partonic description.
This is certainly advantageous when describing scaling of
inelastic nucleon structure functions in the parton model.

The parton model description of diffractive deep inelastic processes in the
Bjorken limit corresponds to the Regge region
($s \approx Q^2 / x \gg Q^2$).
Therefore nucleon DIS at small $x$ can be viewed in terms of virtual-photon
interactions with the Pomeron (${\cal P}$) through its `structure function',
$F_{2{\cal P}}$.
In calculations of $F_{2{\cal P}}$ two contributions are usually included,
an explicit $q\bar q$ component \cite{dola,niza}, and the triple Pomeron
term \cite{niza,kwbd},
$ F_{2{\cal P}}(x_{{\cal P}},Q^2)
= F_{2{\cal P}}^{(q\bar q)}(x_{{\cal P}},Q^2)
+ F_{2{\cal P}}^{(3{\cal P})}(x_{{\cal P}},Q^2)$, where \cite{dola,methoD}
\begin{mathletters}
\label{F2Pom}
\begin{eqnarray}
F_{2{\cal P}}^{(q\bar q)}(x_{{\cal P}},Q^2)
&=& { 4 (10 + 2 \lambda_s)\ \beta_0^2 \over 3 \sigma_{pp} }\
    { N_{sea}\ Q^2 \over Q^2 + Q_0^2 }\
    x_{{\cal P}} (1-x_{{\cal P}})                               \label{Fqq}\\
F_{2{\cal P}}^{(3{\cal P})}(x_{{\cal P}},Q^2)
&=& { g_{3{\cal P}} \over \sqrt{\sigma_{pp}} }
    F_{2N}^{sea}(x_{{\cal P}},Q^2).                             \label{F3P}
\end{eqnarray}
\end{mathletters}
In Eqs.(\ref{F2Pom}) $x_{{\cal P}} = x/y$ is the fraction of the Pomeron's
momentum carried by the struck quark,
$y ( = k\cdot q/p\cdot q = x (1 + M_X^2/Q^2) \approx M_X^2/s)$
is the fraction of the momentum of the nucleon carried by the Pomeron
(four-momentum $k$) and $M_X^2 = (k + q)^2$ is the mass squared of the
produced hadronic debris $X$.
In Eq.(\ref{Fqq}) $\beta_0^2 = 3.4$ GeV$^{-2}$ is the quark---Pomeron
coupling constant \cite{dola}, and $\lambda_s \simeq 0.5$ represents the
weaker coupling of ${\cal P}$ to the strange quark compared with the $u$
and $d$ quarks.
The parameter $N_{sea}$ ($\approx 0.17$ at $Q^2 \sim 4$ GeV$^2$) is determined
by the small-$x$ behaviour of the nucleon sea distribution,
$ x q_{sea}(x \rightarrow 0) \rightarrow
  N_{sea} x^{1 - \alpha_{{\cal P}}(0)} $,
where $\alpha_{\cal P}(0) \approx 1$ is the ${\cal P}$ intercept \cite{dola}.
In Eq.(\ref{F3P}) $g_{3{\cal P}} = 0.364$ mb$^{1/2}$ is the triple Pomeron
coupling constant \cite{3Pexp}.
For the $pp$ total cross section we use the Regge theory-based
parameterisation:\
$\sigma_{pp} = 21.70 s^{\epsilon} + 56.08 s^{-\eta}$ \cite{dlsig}.

For fixed $\nu$, the small-$x$ region also corresponds to small $Q^2$.
At current energies, this means that probing nuclear structure functions at
$x \alt 10^{-3}$ is possible only for $Q^2 \alt 0.5$ GeV$^2$, which is
well below the scaling region.
Since the $Q^2$ dependence of $F_{2{\cal P}}$ is determined by the nucleon
sea quark densities, we need to know how these behave at
low $Q^2$.
Although a rigorous theoretical basis for nucleon structure functions at
very small $Q^2$ is still outstanding, some phenomenological parameterisations
have been constructed \cite{schsj,dl93} by incorporating the
photoproduction limit.
The simplest, and most common method adopted has been to multiply $F_{2N}$
by a factor $\left( Q^2 / (Q^2 + Q_0^2) \right)^{1+\epsilon}$, with the
normalisation and mass parameter $Q_0^2$ fixed by matching the photoproduction
and deep inelastic regions.
In Eq.(\ref{F3P}) we extract $F_{2N}^{sea}$ from the parameterisation
of $F_{2p}$ and $F_{2n}$ in Ref.\cite{dl93}, and for the $q\bar q$ contribution
we include in Eq.(\ref{Fqq}) the factor $Q^2 / (Q^2 + Q_0^2)$.
The parameter $Q_0^2 \approx 0.485$ GeV$^2$, which we take from
Donnachie \& Landshoff \cite{dl93}, may
be interpreted as the inverse size of partons inside the virtual photon.

In diffractive DIS from nuclei Pomeron exchange between the projectile
and two or more constituent nucleons models the interaction of partons
from different nucleons within the nucleus.
The shadowing correction to $F_{2A}$ from ${\cal P}$-exchange
can be written as a convolution of $F_{2{\cal P}}$ and an
exchanged-${\cal P}$ distribution function, $f_{{\cal P}/A}$:
\begin{mathletters}
\begin{eqnarray}
A\ \delta^{({\cal P})} F_{2A}(x,Q^2)
&=& \int_{y_{min}}^A\ dy\
    f_{{\cal P}/A}(y)\ F_{2{\cal P}}(x_{{\cal P}},Q^2)          \label{dFAP}
\end{eqnarray}
where
\begin{eqnarray}
f_{{\cal P}/A}(y)
&=& - { A (A-1)\ 8 \pi \over y }\
    {\rm Re}
    \int_{z' > z} d^2{\bf b}\ dz\ dz'\
    e^{i k_L (z'-z)}                                            \nonumber\\
& & \hspace*{0cm} \times\
    \rho^{(2)}({\bf b},z,z')\ \
    \exp \left( - {A \over 2} \int_z^{z'} {d\zeta \over L_X}
         \right)                                                \label{Apom}
\end{eqnarray}
with $k_L = M\ y$.
\end{mathletters}
The mean free path of the hadronic state $X$ inside the nucleus is
$L_X = (\sigma_{XN}\ \rho({\bf b},\zeta))^{-1}$, and the assumption
is made that the total cross section, $\sigma_{XN}$, for the
state $X$ with a nucleon is independent of the mass $M_X$ \cite{kwbd}
--- in which case we may take $\sigma_{XN} \sim 20-30$ mb, similar to the
vector meson--nucleon cross sections.
The uncertainty in the precise value of $\sigma_{XN}$ is not significant
numerically for small and intermediate $Q^2$, since there
$\delta^{({\cal P})} F_{2A}$ is much smaller than $\delta^{(V)} F_{2A}$.
At higher $Q^2$, however, the larger nucleon sea gives a large
${\cal P}$-exchange contribution, so that above $Q^2 \sim 5$ GeV$^2$
it starts to dominate.

The combined effect of both the VMD and ${\cal P}$-exchange mechanisms is
a total shadowing correction that varies little for $Q^2$ between
$\sim 0.5$ and 10 GeV$^2$.
In this transition region concerns have been expressed about possible double
counting when adding the two components \cite{kwbd}.
Since the VMD contribution is a higher twist effect, this problem could in
principle be averted by keeping only the leading twist piece of
$F_{2{\cal P}}$.
Alternatively, the solution adopted by Kwiecinski \& Badelek \cite{kwbd}
was to exclude from the ${\cal P}$-exchange process those final states $X$
which have a mass comparable to that of the vector mesons
($M_X^2 < M_{X_0}^2 \simeq 1.5$ GeV$^2$), and consequently take as the lower
limit of integration in Eq.(\ref{dFAP}) $y_{min} = x (1 + M_{X_0}^2/Q^2)$.
Numerically, this will be important only at small $x$ ($x \alt 10^{-3}$)
and $Q^2$, while for larger $Q^2$ the separation becomes
redundant because there $y_{min} \rightarrow x$.

In Figs.1--3 we illustrate the effect of shadowing on the ratios of nuclear to
deuteron structure functions (per nucleon) for various nuclei
(with $F_{2A} = (Z/A) F_{2p} + (1-Z/A) F_{2n}
              + \delta^{(V)} F_{2A} + \delta^{({\cal P})} F_{2A}$).
For $F_{2p}$ and $F_{2n}$ we use the parameterisations of
Donnachie \& Landshoff \cite{dl93}.
(In Ref.\cite{dl93} the neutron structure function was extracted from $F_{2D}$
without taking into account the possible shadowing corrections in the
deuteron. However, these are much smaller than the shadowing corrections
for the heavy nuclei considered here and would have little
effect on the final $A/D$ ratios.)

Figure 1 shows the prediction of the model for the $Xe / D$
structure function ratio, compared with the recent E665 data \cite{e665Xe}.
The dashed curve is the result of the VMD model, while the solid curve
includes in addition the ${\cal P}$-exchange component.
Clearly the bulk of the shadowing is due to the rescattering of vector mesons.
Note that the data (and the curves) at each $x$ correspond to a different
average value of $Q^2$, ranging from $\langle Q^2 \rangle \approx 0.03$ GeV$^2$
at the lowest $x$ value to $\langle Q^2 \rangle \approx 10$ GeV$^2$
at $x \approx 0.1$.
The flattening of the curves as $x \rightarrow 0$ is a direct consequence
of the vanishing (like $\sim Q^2$) of the total $F_{2A}$ as $Q^2 \rightarrow 0$
(i.e. a smooth extrapolation to the photoproduction limit).
As expected, the role of two-particle correlations is found to be rather
small, with the result of using an independent particle approximation
differing only by some $5-10\%$ from that in Fig.1.
Of greater significance are the effects due to projectile attenuation, which
reduce the amount of shadowing by up to $30-40\%$.
Indeed, for larger nuclei the attenuation is even stronger, so that eventually
saturation with $A$ is reached.
This is illustrated in Fig.2 where we plot $F_{2A}/F_{2D}$ for $\nu = 150$
and 250 GeV (at $x = 0.0025$), and compare with the recent Fermilab data
for $Ca, Xe$ and $Pb$, taken at $0.0018 < x < 0.0032$
\cite{e665prel}.

For completeness, in Fig.3 we also compare the model predictions
with the high-precision NMC data \cite{nmc} on lighter nuclei
($Ca$ and $C$), taken at $0.5 \alt Q^2 \alt 10$ GeV$^2$.
Clearly the agreement with the data is quite good for both
$Ca$ and $C$.


Having obtained good agreement with the data on heavy nuclei, we next calculate
the shadowing corrections to the deuteron structure function in the region
$x \agt 10^{-5}$ and $Q^2 \agt 0.03$ GeV$^2$, where the recent Fermilab data
were collected \cite{e665D}.
Traditionally in DIS on the deuteron, in which the proton and neutron
are held together very weakly, nuclear effects have been ignored.
Recent calculations \cite{Dshad,methoD}, however, have suggested small,
but possibly non-negligible corrections at $Q^2 \sim 4$ GeV$^2$.

For $\gamma^* D$ scattering the Glauber formalism involves just the
double scattering term in the multiple scattering series.
In this case the shadowing corrections to $F_{2D}$ are given by
Eqs.(\ref{dFAV}) and (\ref{dFAP}), except now the distribution of nucleons
in the deuteron is given by the deuteron form factor \cite{methoD},
$  S_D({\bf k}^2)
 = \int_0^{\infty} dr \left( u^2(r) + w^2(r) \right) j_0(|{\bf k}| r), $
where $u(r), w(r)$ are the $S,D$-wave deuteron wavefunctions taken from
realistic $NN$ potential models \cite{NNmodel}.

Apart from the VMD and ${\cal P}$-exchange contributions,
another potential source of shadowing in $D$ is the exchange of (unreggeised)
mesons, which was first shown by Kaptari et al. \cite{kaptar} to give sizeable
antishadowing corrections to $F_{2D}$ at $Q^2 \sim 4$ GeV$^2$.
In the non-relativistic approximation, this contribution is given by
the convolution of an exchanged-meson distribution $f_{M/D}(y)$
\cite{kaptar,methoD}, with the meson structure function $F_{2M}(x/y,Q^2)$.
Unfortunately, experimentally only $F_{2\pi}$ is known, and then only in
the region $x \agt 0.2$ and $Q^2 \agt 4$ GeV$^2$.
In addition there are no data on pion photoproduction which could constrain
the $Q^2 \rightarrow 0$ extrapolation parameters for $F_{2\pi}$.
As a simple solution we take the (real) pion structure function measured
in Drell-Yan reactions (as parameterised in Ref.\cite{mrspi},
with the fit that corresponds to a 20\% pion sea at $Q^2 = 5$ GeV$^2$),
and apply a phenomenological factor
$\left( Q^2 / (Q^2 + Q_0^2) \right)^{1+\epsilon}$,
motivated by Regge theory.
For $Q_0^2$ we use the same value as in the analysis of the $\gamma p$ data
(see above), since $Q_0^2$ should be independent of the target hadron if it is
to be interpreted as the `mass' of the $q\bar q$ pair in the photon.
Although the extrapolation of $F_{2M}$ is the main uncertainty in the
calculation, fortunately the meson-exchange contribution turns out to be
small in the region of $x$ and $Q^2$ covered by the E665 data
(its contribution at $x \sim 10^{-3} (10^{-2})$ is some 6\% (13\%) of the
total shadowing correction to $F_{2D}$).
This result is largely independent of the model $D$ wavefunctions in
Ref.\cite{NNmodel} or the meson--nucleon form factor used.

In Fig.4 we show the ratio of deuteron to proton structure functions
with (solid curve) and without (dotted curve) the VMD, ${\cal P}$- and
meson-exchange shadowing corrections, and compare with the recent data
from the E665 Collaboration \cite{e665D}.
For each $x$ the E665 data were taken at various photon energies, ranging
from $\nu \sim 300$ GeV at the lowest $x$ values down to $\nu \sim 140$ GeV
at $x \approx 10^{-2}$.
Also shown in Fig.4 are the NMC data at $Q^2 = 4$ GeV$^2$, and the
corresponding theoretical curves (for $x > 0.05$).
The net effect of shadowing is a decrease in $F_{2D}/F_{2p}$ by less than
$\sim 3\%$ over the experimental range of $x$ and $Q^2$.
As a fraction of the total $F_{2D}$ this amounts to $\alt 2-3\%$.
Consequently, for $\nu \sim 150 - 350$ GeV, the free neutron structure function
is $\sim 5\ (2)\%$ larger at $x \sim 10^{-4}\ (10^{-2})$ than the bound neutron
structure function, extracted from the deuteron data without account of
shadowing.
Unfortunately the E665 data at present do not distinguish between the
shadowing and no-shadowing scenarios, and it is essential that the errors
be reduced before any unambiguous conclusions can be reached.
As an independent check, it would be useful if the statistics and range
of $x$ of the neutrino--proton DIS data were also improved.
This would allow the individual quark flavour distributions in the proton
to be determined, and the neutron structure function inferred by
charge symmetry.

In conclusion, we have seen that the two-phase VMD/Pomeron-exchange model
can give a quantitative description of the shadowing observed by the E665
and NM Collaborations in nuclear to deuterium structure function ratios
at low $x$ and $Q^2$.
The apparent saturation of shadowing at $x \alt 10^{-3}$ in $Xe/D$
is naturally reproduced by a careful extrapolation to the total structure
function to the photoproduction limit.
Furthermore, saturation with increasing $A$ is found to be largely a
consequence of the attenuation of the projectile debris inside the nucleus.
The good agreement with the present nuclear data should also give us some
confidence in the predictions for shadowing in the deuteron itself.

\vspace*{1cm}


\parindent 0cm
We would like to thank A.Donnachie and W.J.Stirling for providing their
latest structure function parameterisations, and N.N.Nikolaev and A.Yu.Umnikov
for helpful communications.
This work was supported by the Australian Research Council.


\newpage
{\bf Figure Captions.}

{\bf 1}. Shadowing in the $Xe$ nucleus.
         The dashed curve is the result of the VMD model, while the solid
         curve includes in addition the Pomeron-exchange contribution.
         The data are from the E665 Collaboration \protect\cite{e665Xe},
         and both the statistical and systematic errors are shown.      \\

{\bf 2}. Shadowing as a function of mass number $A$.
         The theoretical curves are evaluated at $x=0.0025$, for
         $\nu$ = 150 GeV (dashed) and $\nu$ = 250 GeV (solid),
         and compared with the preliminary E665 data for
         $Ca, Xe$ and $Pb$, taken for $0.0018 < x < 0.0032$
         \protect\cite{e665prel}.       \\

{\bf 3}. Shadowing in the $Ca$ and $C$ nuclei.
         The data are from the NMC \protect\cite{nmc},
         with statistical and systematic errors added in quadrature.    \\

{\bf 4}. Ratio of deuteron to proton structure functions
         with (solid curve) and without (dotted) shadowing corrections.
         The data (full circles) are from the E665 Collaboration
         (with statistical and systematic errors shown separately),
         and from the NMC (open circles) which were taken
         at $Q^2 = 4$ GeV$^2$.          \\

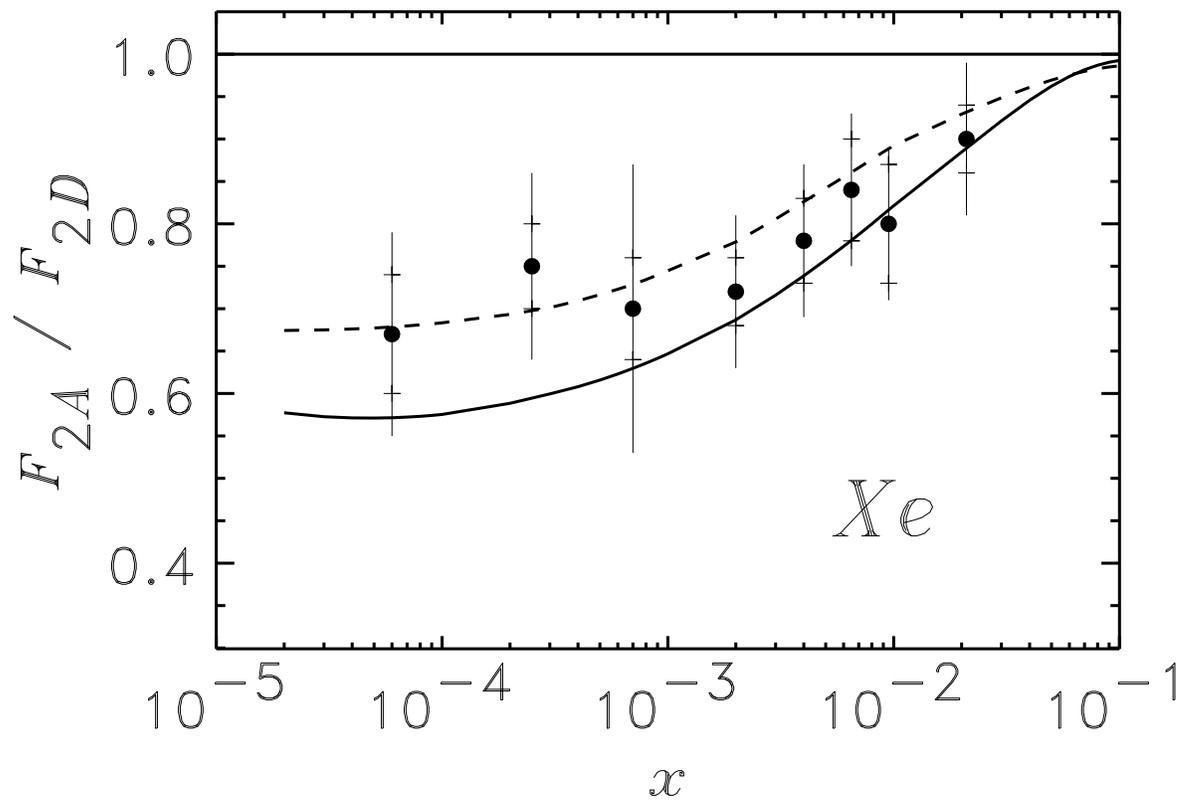
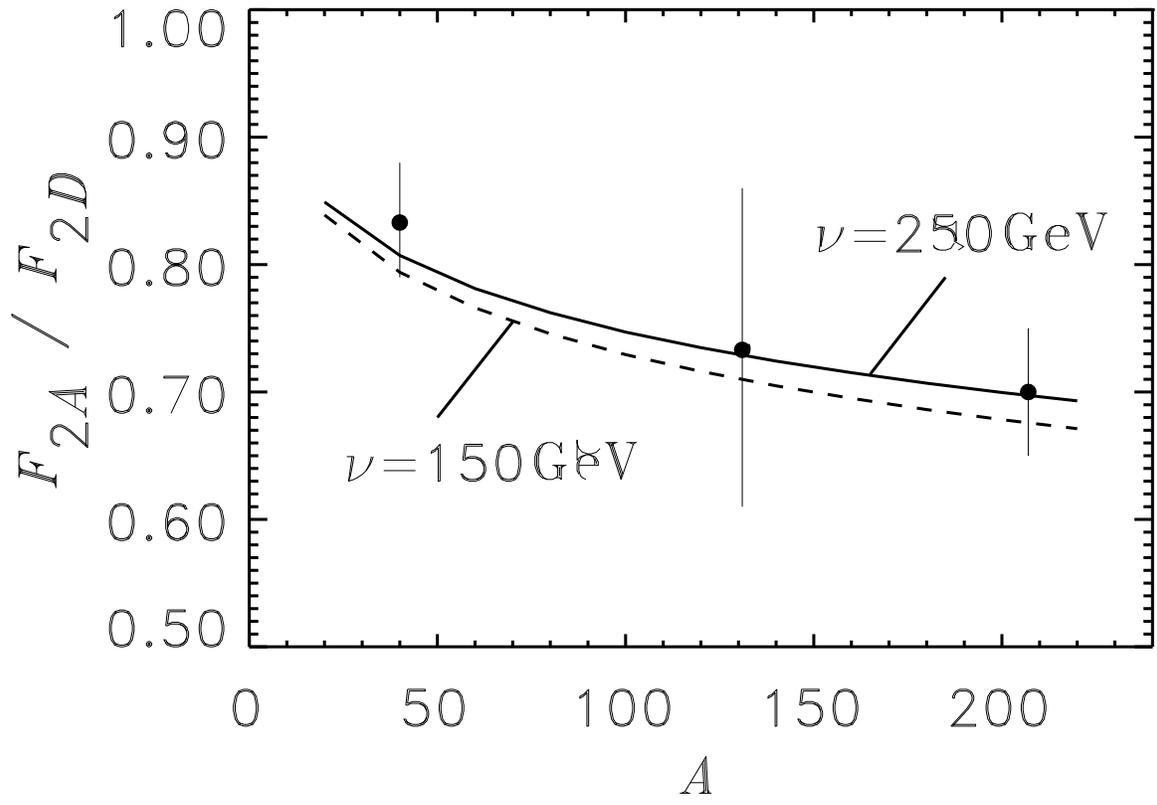
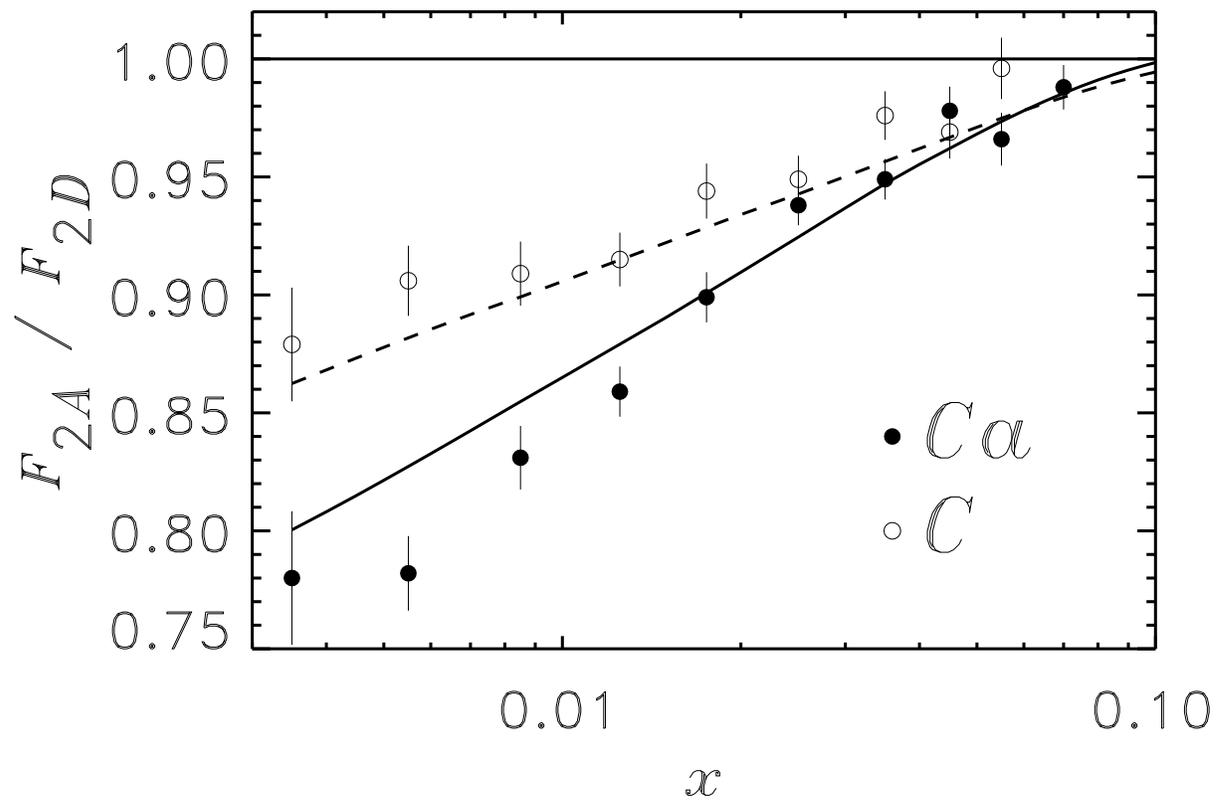
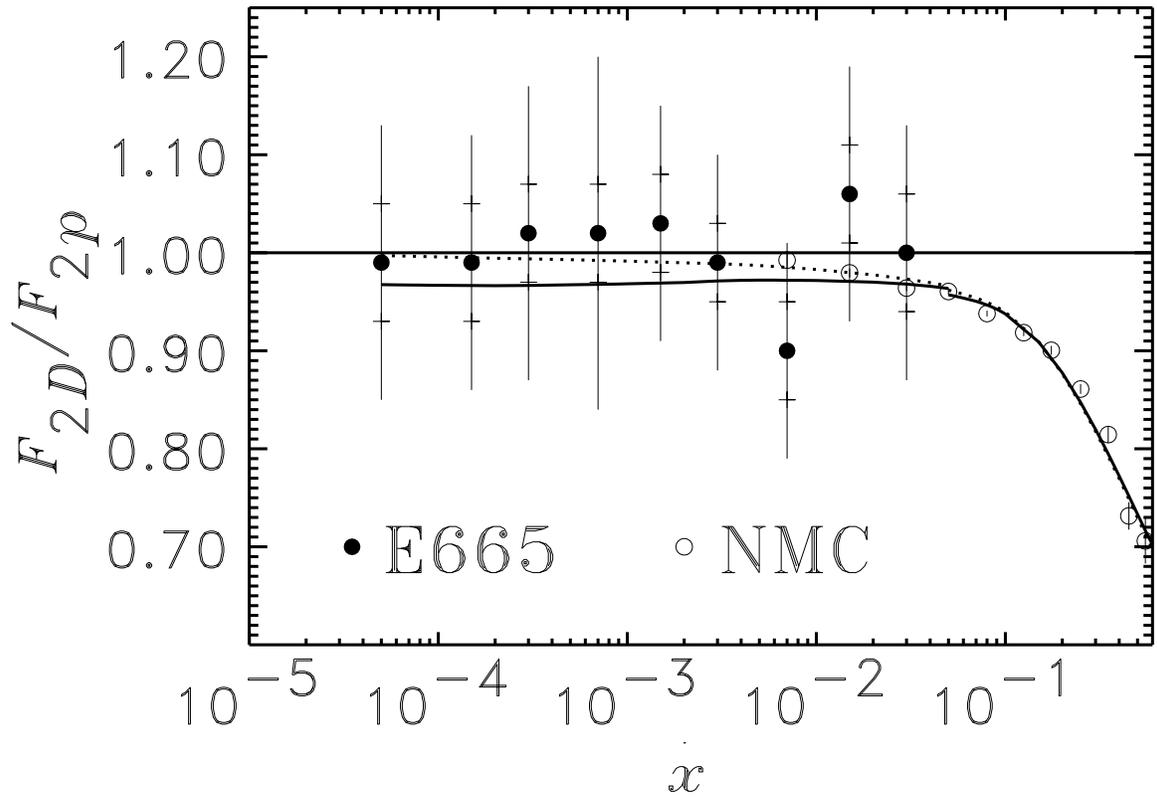
\end{document}